\documentclass[]{aa}
\usepackage{natbib}
\usepackage{epsfig}
\def\ut#1{\mathop{\vtop{\ialign{##\crcr
     $\hfil\displaystyle{#1}\hfil$\crcr\noalign
     {\kern1pt\nointerlineskip}\hbox{$\hfil\sim\hfil$}\crcr
     \noalign{\kern1pt}}}}}

\def\undersymbol#1#2{\mathop{\vtop{\ialign{##\crcr
     $\hfil\displaystyle{#2}\hfil$\crcr\noalign
     {\kern1pt\nointerlineskip}\hbox{$\hfil#1\hfil$}\crcr
     \noalign{\kern1pt}}}}}

\def\degr{^0}

\begin{document}
\title{A super massive black hole binary in 3C66B: future observational perspectives}
\author{F. De Paolis, G. Ingrosso, A. A. Nucita} \offprints{depaolis@le.infn.it} \institute{ Dipartimento
di Fisica, Universit\`a di Lecce, and {\it INFN}, Sezione di
Lecce, Via Arnesano, CP 193, I-73100 Lecce, Italy}
\authorrunning{De Paolis et al.}
\titlerunning{A SMBH binary in 3C66B: future observational perspectives}

\abstract{Supermassive black hole binaries (SMBHBs) may exist in
the centers of galaxies and active galactic nuclei (AGN) and are
expected to be fairly common in the Universe as a consequence of
merging processes between galaxies. The existence of SMBHBs can be
probed by looking for double nuclei in galaxy centers or, more
easily, detecting periodic behavior in the observed radio light
curves. In a recent paper, Sudou et al. \cite{sudou2003} announced
the first direct observation of an SMBHB. Using VLBI observations
they found that the unresolved radio core of the radio galaxy
3C66B shows a well defined elliptical motion with a period of
$1.05\pm 0.03$ yrs, implying the presence of a couple of massive
black holes in the center of the galaxy. In the present paper we
study the astrophysical implications of the existence of such an
SMBHB in 3C66B. In particular we focus on the information that can
be obtained from detecting a signal periodicity either in the
$X$-ray and/or $\gamma$-ray light curves as a consequence of the
motion of the black holes. These observations could be used to
extract further information on the physical parameters of the
SMBHB and partially solve the system parameter degeneracy. The
detection of the gravitational wave spectrum emitted by such
system may be used to completely determine the physical parameters
of the binary system. }
\maketitle

\section{Introduction}

It is now well established that galaxies generally contain massive
black holes in their nuclei (Rees \citeyear{rees}, Kormendy \&
Richstone \citeyear{kr}, Richstone et al. \citeyear{rab}) and
since merging events are expected to be rather common  (see e.g.
White \citeyear{white})  supermassive black hole binaries
(hereinafter SMBHBs) may be found frequently in galactic cores
(Begelman et al. \citeyear{bbr}).

Up to now, several astrophysical phenomena have been attributed to
SMBHBs. In fact, their presence influences the stars in the galaxy
center regions whose mass density profile, as observed by
numerical simulations (Nakano \& Makino \citeyear{nakanomakino}),
may show a $\rho \propto r^{-1/2}$ profile typical of giant
galaxies that have experienced many merging processes in their
lifetime.

Moreover, if the black holes are rotating and have misaligned
spins, plasma beams (aligned with the black hole axis) precess
with period of $10^3-10^4$ yr and curved jets should be observed.
This phenomenon happens in double-double radio galaxies
(Schoenmakers \citeyear{schoenmakers2000}) and in both S and
X-shaped galaxies (Leahy \& Williams \citeyear{leahy1884}, Parma
et al. \citeyear{parma1985}, Wang et al. \citeyear{wang2003}).

Periodic outburst activity, as for example in the quasar OJ $287$
(see e.g. Sillanp$\ddot{a}\ddot{a}$ et al. \citeyear{shv}, Lehto
\& Valtonen \citeyear{lv}) and periodic behavior in the radio,
optical, $X$-ray and $\gamma$-ray light curves of many AGNs are
also possibly related to the presence of a central massive binary
black hole creating a jet either aligned along the line of sight
or interacting with an accretion disk (Yu \citeyear{yu}).

Up to now, the search for $X$-ray and/or $\gamma$-ray variability
was considered as a method to probe the existence of an SMBHB in
the center of a galaxy. For example, at least three Mkn objects
(Mkn 501, Mkn 421 and Mkn 766) reveal periodic behavior in the
observed signal so that in recent studies they have been
considered as possibly containing SMBHBs (see Rieger \& Mannheim
\citeyear{rm}, De Paolis et al. \citeyear{depaolismkn1} and De
Paolis et al. \citeyear{depaolismkn2}).

However, the best tracer of the existence of SMBHB in a galaxy
core remains the detection of the Keplerian orbital motion of some
emission component close to the two black holes and, eventually,
the observation of eclipse episodes as in the case of the AGN OJ
287 (Sadun et al. \citeyear{sadun}, Lehto \& Valtonen
\citeyear{lv}, Pietila \citeyear{pietila1998}).

Unfortunately, a technical problem is that the typical black hole
separation ($\sim 10^{16}-10^{17}$ cm), for a host galaxy located
at the distance of $\simeq 100$ Mpc, corresponds to a separation
of $\sim 10-100$ $\mu$arcsec and only recently has the required
accuracy in position measurements become available by the
phase-referencing very long baseline interferometry technique at
radio frequencies.

In a recent paper Sudou et al. (\citeyear{sudou2003}), using the
VLBI facility, announced the first detection of a well defined
Keplerian orbital motion of the radio emission component in a pair
of well separated radio sources, 3C66B and 3C66A. The natural
explanation of this observation is that the radio galaxy 3C66B
hosts an SMBHB at its center.

On the assumption that the SMBHB orbit is circular, Sudou et al.
(\citeyear{sudou2003}) have estimated, by a best fitting
procedure, the physical parameters of the system and found a mass
density value of $2\times 10^{15}$ M$_{\odot}$ pc $^{-3}$, larger
than of the massive object in NGC4258 and SgrA$^*$.

However, radio observations do not make it possible to determine
all the binary system parameters since the results depend on the
mass ratio $q$ between the two components, on the eccentricity of
the system, and on the semi-major axis.

The aim of this paper is to present a complete analysis of the
SMBHB in question by relaxing the assumption of a circular
orbit\footnote{It is expected in fact that if SMBHBs come from
merging events between galaxies they could be in elliptical orbits
with eccentricities up to $0.8-0.9$ (Fitchett
\citeyear{fitchett}).} and indicating some direction for further
observations which may solve the parameter degeneracy for the
system in 3C66B.

In fact, a galaxy nucleus hosting an SMBHB of which one of the
components is emitting a jet with Lorentz factor $\gamma _{\rm L}$
towards the observer may show periodic light curves both at
$X$-ray and $\gamma$-ray wavelengths (see Rieger \& Mannheim
\citeyear{rm} and De Paolis et al. \citeyear{depaolismkn1}).

Thus, assuming that the radio jet in 3C66B is emitted by the less
massive black hole (with mass $m$) which is moving in an
elliptical orbit (with semi-major axis $a$ given by the VLBI radio
observations, and generic eccentricity $e$) around its companion
(having mass $M$), we study both the expected $X$-ray and
$\gamma$-ray signal periodicity $P_{\rm obs}$ and the flux ratio
$f$ between maximum and minimum signal as a function of the jet
Lorentz factor $\gamma _{\rm L}$ and of the power law spectral
index $\alpha$ of the photon flux.

We show how the measurements of $f$ and $P_{\rm obs}$ may be used
to extract further information about the physical parameters of
the SMBHB system hosted by the radio galaxy 3C66B.

We also consider the gravitational wave spectrum and amplitude
emitted by the SMBHB as a function of the orbital parameters of
the system and compare the expected signal with the capabilities
of the next generation space-based interferometers LISA and
ASTROD.

The paper is structured as follows: in Sect. 2 we review the main
properties of the radio galaxy 3C66B and in particular the
observed parameters of the SMBHB at its center. In Sect. 3 we
discuss the expected periodicity of the light curves in the
$X$-ray and/or $\gamma$-ray band and show how the flux ratio $f$
depends on the system parameters. We show how one can use the
expected periodicity $P_{\rm obs}$ and $f$ to constrain the
physical parameters of the binary system. We have performed this
analysis by assuming that the jet inclination angle to the line of
sight is $i\simeq 1/{\gamma_{\rm L}}$ as expected for typical
blazars. In addition, we have considered also the more realistic
case in which $i$ may be estimated by radio observations. However,
in both cases the SMBHB parameters may be completely determined
only if the spectrum of the gravitational waves emitted by the
system is detected. Finally, our conclusions are given in Sect. 4.

\section{An SMBHB in the radio galaxy 3C66B}

Using the very-long-baseline-interferometer, Sudou et al.
(\citeyear{sudou2003}) recently announced the first direct
evidence of the existence of an SMBHB in a distant galaxy.

They looked for the Keplerian orbital motion of a radio emission
component in a pair of well separated ($\simeq$ 6 arcmin) radio
sources, 3C66B and 3C66A; 3C66B is a radio galaxy at redshift
$z=0.0215$ (Matthews et al. \citeyear{matthews1964}) whereas 3C66A
is a more distant BL Lac object ($z=0.44$, Miller
\citeyear{miller1978}) which was used as the stationary position
reference to 3C66B. In fact, observing both objects at 2.3 and 8.4
GHz (Sudou et al. \citeyear{sudou2003}) the radio core position of
3C66B at each epoch was measured with respect to 3C66A. This
revealed a time variation of the 3C66B core position which is well
fitted by an elliptical motion with an averaged period estimated
to be $1.05\pm 0.03$ yrs (see Table \ref{table1} and Sudou et al.
\citeyear{sudou2003} for more details).

Since the radio core is located at the root of the jet where the
optical depth for synchrotron self-absorption is of the order of
unity (Lobanov \citeyear{lobanov}), it is expected that the 2.3
GHz core is located at a greater distance from the central engine
than the 8.4 GHz core. Thus, the observation that the 2.3 GHz
semi-major axis ($243 \pm 30$ $\mu$arcsecs) is larger than the 8.4
GHz one ($45 \pm 4$ $\mu$arcsecs) is a proof that the jet in 3C66B
also suffers a precessional motion. According to Sudou et al.
\citeyear{sudou2003}, the most plausible explanation for this
result is that while the 2.3 GHz core motion maps the jet
precession, the elliptic motion observed in the 8.4 GHz band well
describes the orbital motion of an SMBHB in 3C66B with a Keplerian
period $P_{K} =1.05\pm 0.03$ yrs.

\begin{figure}[htbp]
\begin{center}
\vspace{7.5cm} \includegraphics{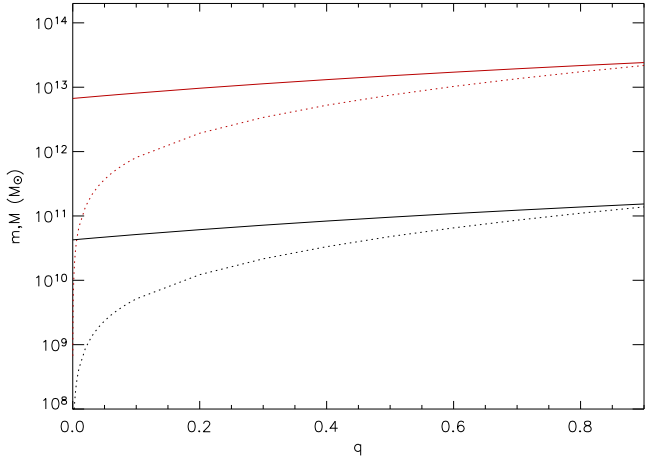} \caption{The masses of the two black holes
$m$ (dashed line) and $M$ (solid line) are given as a function of
the mass ratio $q$. Red lines correspond to the orbital maximum
semi-major axis $a\simeq 5.4\times 10^{16} (1+q)$ cm at 8.4 GHz
while black lines correspond to $a\simeq 2.9\times 10^{17} (1+q)$
cm at 2.3 GHz.} \label{figure1}
\end{center}
\end{figure}

Since, as stated above, observations at 8.4 GHz are believed to
reflect the radio core motion of the SMBHB more closely than those
at 2.3 GHz, assuming that the jet is emanating from the less
massive black hole (of mass $m<M$), Sudou et al.
\citeyear{sudou2003} estimated the semi-major axis $a$ of the
fitted orbit to be the upper limit of the orbital radius of the
SMBHB, i.e.
\begin{equation}
a_{max}\simeq 5.4\times 10^{16} (1+q)~~{\rm cm}, \label{uppersep}
\end{equation}
where $q=m/M$ is the mass ratio between the two binary components.
With this assumption, and from the definition of the Keplerian
period
\begin{equation}
P_{K}=2\pi \sqrt{\frac{r^3}{G(M+m)}}~, \label{periodo}
\end{equation}
the following relation holds
\begin{equation}
M \simeq 4.6 \times 10^{10} (1+q)^2~~{\rm M_{\odot}}~,
\label{set1}
\end{equation}
which makes it possible to estimate the masses of the two
components once the mass ratio $q$ is given. In Fig. \ref{figure1}
the two black hole masses $m$ and $M$ of the black holes are shown
as a function of the mass ratio $q$.

\begin{table}
\caption{Parameters of the fitted elliptical motion of the radio
core in 3C 66B monitored in the two radio bands 2.3 and 8.4 GHz.
Data are taken from Sudou et al. \citeyear{sudou2003}}
\begin{center}
\begin{tabular}{|c|c|c|}
\hline
Radio Bands                   & 2.3 GHZ         &8.4 GHZ          \\
\hline
Semi-major axis ($\mu$arcsec) &  243 $\pm$ 30   &   45  $\pm$ 4   \\
Axial ratio              &  0.31$\pm$ 0.17 &  0.24 $\pm$ 0.14\\
Period (yrs)             &  1.10 $\pm$ 0.06&  1.02 $\pm$ 0.04\\
\hline
\end{tabular}
\end{center}
\label{table1}
\end{table}

As is clear from Table \ref{table1}, the estimated orbit
semi-major axis $a$ is different at different wavelengths. In
fact, it is $a_{max}\simeq 5.4\times 10^{16}(1+q)$ cm at 8.4 GHz
and increases by a factor $\eta\simeq 5.4$ at 2.3 GHz becoming
$a_{max}\simeq 2.9\times 10^{17}(1+q)$ cm. Thus, the mass $M$ in
the previous equation is increased by a factor $\eta^3$ (as is
also clearly stated in Sudou et al. \citeyear{sudou2003}) if the
semi-major axis is taken to be given by the observations at $2.3$
GHz.

Here we would like to emphasize that observations at higher
frequencies should allow to map more accurately the core motion in
3C66B giving a better estimate of the orbit semi-major axis which
is expected to be smaller than $a_{max}\simeq 5.4\times
10^{16}(1+q)$ cm.

Sudou et al. (\citeyear{sudou2003}) assumed that the binary system
is characterized by a circular orbit (i.e. $e = 0$). However, if
binary black holes originated from galactic mergers, they could be
in eccentric orbits, and eccentricity values up to $0.8$-$0.9$ are
not necessarily extreme (Fitchett \citeyear{fitchett}). Of course,
due to gravitational wave emission, orbits tend to circularize but
this happens within a time-scale of the same order of magnitude as
the merging time-scale (Peters \citeyear{peters}, Fitchett
\citeyear{fitchett}). Therefore, if an SMBHB is found at the
center of a galaxy, it may happen that the components are still in
eccentric orbits. Consequently, the axial ratio quoted in Table
\ref{table1} gives an upper limit to the true orbital eccentricity
$e$ since the binary orbit is seen projected on the plane of the
sky. Thus, the possible SMBHB in 3C66B may be characterized by an
orbit with with an eccentricity up to $e_{max}\simeq 0.96$.
Relaxing the assumption of circular orbits will produce major
changes in the estimation of both the gravitational wave amplitude
and the coalescing time-scale of the SMBHB possibly hosted by
3C66B.

\section{Signal periodicity in the $X$-ray and $\gamma$-ray light curves and
the determination of the maximum-to-minimum flux ratio}

Periodic outburst activities and periodic behavior in the radio,
optical, $X$-ray and $\gamma$-ray light curves of many AGNs are
believed to be related to the presence of an SMBHB in the center
of the host galaxy (Yu \citeyear{yu}).

On this basis, Rieger \& Mannheim (\citeyear{rm}) (but see also De
Paolis et al. \citeyear{depaolismkn1} for a generalization to
orbits with generic eccentricity values) have recently proposed a
method to determine the orbital parameters of the binary system
from the observed quantities, i.e. the signal periodicity $P_{\rm
obs}$, the flux ratio $f$ between maximum and minimum signal and
the power law spectral index $\alpha$ of the photon flux.

According to these models, the periodicity in the flaring state is
assumed to be the consequence of the orbital motion of a
relativistic jet in the binary black hole. Therefore, the observed
signal periodicity has a geometrical origin due to
Doppler-shifted modulation.

As a consequence of the binary orbital revolution around the
center of mass, the expected signal period $P_{\rm obs}$ is
related to the Keplerian period $P_{\rm K}$ by (Rieger \& Mannheim
\citeyear{rm})
\begin{equation}
P_{\rm obs}= (1+z) \left(1- \frac{v_z}{c}\cos i \right)P_{\rm K}~,
\label{observedperiodicity}
\end{equation}
where, according to the standard model for the jet bulk motion
(see e.g. Spada \citeyear{spada}), the jet outflow velocity $v_z$
is given by
\begin{equation}
v_z\simeq c(1-1/\gamma_{\rm L}^2)^{1/2}~, \label{jetvelocity}
\end{equation}
and $i$ is the inclination angle between the jet axis and the line
of sight. In the case of blazars, it is usually assumed that
$i\simeq 1/\gamma _{\rm L}$ (Robson \citeyear{robson}, Urry and
Padovani \citeyear{urrypadovani1995}). However, in the case of
radio galaxies this assumption may not always hold and $i$ (or at
least an upper limit) can be estimated by direct observations. In
the following we develop the necessary formalism in general
(without assumptions for the inclination angle) and then we will
apply it to the radio galaxy 3C66B by assuming either $i\simeq
1/\gamma_{\rm L}$ or the available radio observations which
restrict the parameter range (Hardcastle et al.
\citeyear{hardcastle1996}).

\begin{figure}[htbp]
\begin{center}
\vspace{7.0cm} \includegraphics{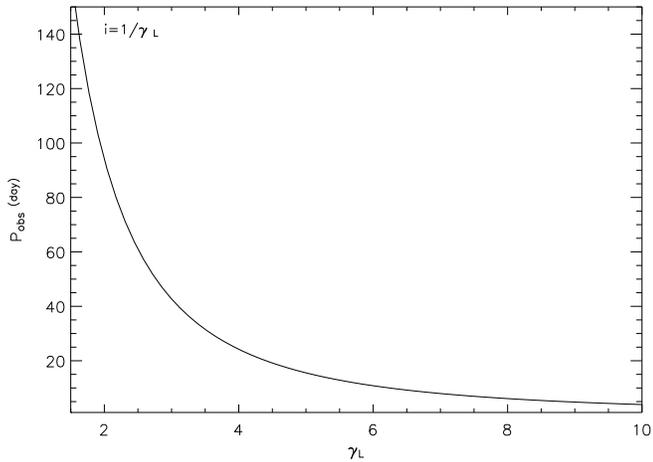} \caption{The expected light curve
periodicity as a function of the jet Lorentz factor $\gamma_{\rm
L}$ is shown assuming the relation $i\simeq 1/\gamma _{\rm L}$.}
\label{figure2}
\end{center}
\end{figure}

For a typical blazar in which $i\simeq 1/\gamma _{\rm L}$, we show
in Fig. \ref{figure2} the dependence of the expected light curve
periodicity on $\gamma_{\rm L}$.

Thus, for the 3C66B galaxy it is expected that in the $X$-ray
and/or $\gamma$-ray band the signal periodicity $P_{\rm obs}$
corresponding to $P_{\rm K}\simeq 1.05$ yrs is in the range $\sim
10-150$ days as a function of $\gamma_{\rm L}$ in the case of
$i\simeq 1/\gamma _{\rm L}$.
3C66B is a large source (several arcminutes in the direction of
largest angular size) and is characterized by a jet and a
counterjet with a high degree of asymmetry in the surface
brightness up to distances of a few kpc from the radio nucleus,
becoming more symmetric on largest scales.

The flux asymmetry may be explained by considering relativistic
beaming, i.e. the whole jet material moves at the same bulk
velocity $\beta c$ and the jet and counterjet make the same angle
$i$ to the line of sight. In this case, the emission is
Doppler-beamed towards or away from the observer and the ratio $R$
between the jet and counterjet radio emission is given by
(Hardcastle et al. \citeyear{hardcastle1996})
\begin{equation}
R=\left(\frac{1+\beta \cos i}{1-\beta \cos i}\right)^{2+\alpha
_r}~,
\end{equation}
where $\alpha _r$ is the spectral index in the radio band.
Measuring the integrated flux ratios of the jet and counterjet at
different angular distances, it is possible to find $\beta \cos i$
as a function of distance from the nucleus. This procedure
constrains $\beta \cos i$ to be in the range $0.25-0.6$. Moreover,
values of $\beta \cos i$ around $0.6$ constrain the angle to the
line of sight of the emerging jet to be $i\ut< 53\degr$.
\begin{figure}[htbp]
\begin{center}
\vspace{7.0cm} \includegraphics{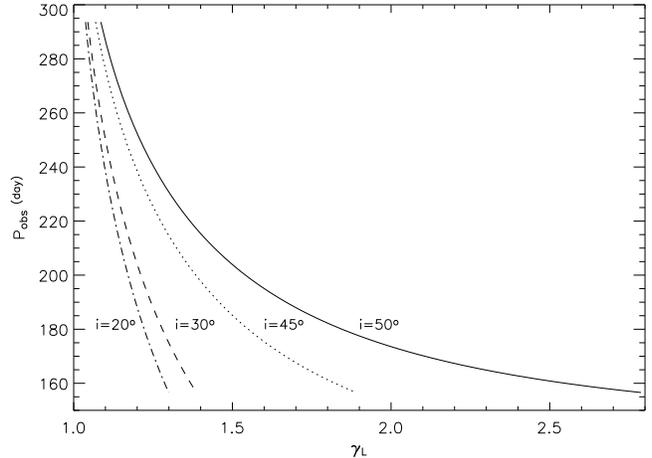} \caption{The expected light curve
periodicity as a function of $\gamma_{\rm L}$ is shown for
different jet viewing angles $i$.}
\label{periodicitadiversiangoli}
\end{center}
\end{figure}
For our purposes, leaving the jet inclination angle $i$ as a free
parameter for the bulk motion velocity $\beta$ one has the obvious
condition $\beta _{min}<\beta<\beta _{max}$, with $\beta
_{min}=0.25/\cos i$ and  $\beta _{max}=0.6/\cos i$. Consequently,
the Lorentz factor is in the range $(1-\beta
_{min}^2)^{-0.5}<\gamma _{\rm L}< (1-\beta _{max}^2)^{-0.5}$.

In this way we can evaluate the expected $X$-ray and/or
$\gamma$-ray light curve periodicity as a function of $\gamma_{\rm
L}$ for different jet inclination angles $i$ by using Eq.
(\ref{observedperiodicity}). In Fig.
\ref{periodicitadiversiangoli}, we plot $P _{obs}$ for the angles
$i=20\degr,~30\degr,~45\degr$ and $50\degr$. Note that the
observed periodicity is always less than one year.

Let us further assume that the jet is emitted by the less massive
black hole (of mass $m$) and that, as usual, the jet axis is
parallel to the orbital angular momentum vector. \footnote{We
would like to mention that a generalization of our treatment to
the more general case with arbitrary direction of the jet axis
with respect to the orbital angular momentum vector requires the
introduction of a set of quantities (the position angles of the
jet in the sky plane, the jet precession opening half-angle, the
inclination angle of the orbital plane, etc.) following, for
example, a similar approach to that proposed by Abraham \& Carrara
(\citeyear{carrara}). However, this approach requires the
knowledge of a number of parameters that are only weakly
constrained by the observations. A comprehensive treatment of this
case will be presented elsewhere.} Moreover, we assume that the
non-thermal $X$-ray and/or $\gamma$-ray radiation propagates
outwards from the core along the jet with Lorentz factor $\gamma
_{\rm L}$. In this way, the observed flux modulation due to
Doppler boosting can be written as (Rieger \& Mannheim
\citeyear{rm})

\begin{equation}
S(\nu) = \delta ^{3+\alpha}S^{\prime}(\nu)~,
\label{flux}
\end{equation}
where $\alpha$ is the source spectral index \footnote{In the
following,  typical values of the power law index $\alpha$ are
assumed to be in the range 1-2 (for details see Guainazzi,  et al.
\citeyear{gvm} and Boller et al. \citeyear{bkt}).} and the Doppler
factor is given by
\begin{equation}
\delta = \frac{\sqrt{1-(v_z^2+ v_{ls}(\theta)^2)/c^2}}{1-(v_z \cos
i + v_{ls}(\theta) \sin i )/c}~. \label{doppler}
\end{equation}
Here, $v_{ls}$ is the component of the less massive black hole
velocity along the line of sight and $\theta$ the polar angle
defining the position of $m$ in its orbit.

As discussed in De Paolis et al. (\citeyear{depaolismkn1}), the
maximum and minimum values of the velocity $v_{ls}$ are attained
for $\theta = \pi /2$ and $\theta = 3\pi /2$, respectively,
corresponding, through Eq. (\ref{doppler}), to the two possible
values $\delta_{max}$ and $\delta_{min}$ of the Doppler factor.
With the assumption that the periodicity in the observed signal is
due to the orbital motion of the binary black hole, one obtains
from Eq. (\ref{flux}) the condition
\begin{equation}
\delta _{max}/ \delta_{min} \simeq f^{1/(3+\alpha)}~,
\label{dopplerratio}
\end{equation}
where $f$ is the observed maximum to minimum flux ratio. Once the
value of $v_{ls}$ is known the previous equations give (for
details see De Paolis et al. \citeyear{depaolismkn1})
\begin{equation}
\begin{array}{l}
\displaystyle{\frac{M}{(m+M)^{2/3}}=
\frac{P_{\rm obs}^{1/3}}{[2\pi(1+z)G]^{1/3}}\frac{c}{\sin i}}\times \\ \\
~~~~~~~~~~~~~~~~\displaystyle{
\frac{f^{1/3+\alpha}-1}{f^{1/3+\alpha}+1}\left(1
-\frac{v_z}{c}\cos i \right)^{2/3}} (1-e^2)^{1/2}~.
\end{array}
\label{massratio1}
\end{equation}

VLBI observations may be used to constrain the masses of the two
black holes in 3C66B so that the previous equation can be solved
with respect to $f$ giving
\begin{equation}
f= \left[\frac{1+k}{1-k}\right]^{3+\alpha}~, \label{fluxratio}
\end{equation}
with
\begin{equation}
\begin{array}{l}
\displaystyle{k =
\frac{M}{(m+M)^{2/3}}\frac{[2\pi(1+z)G]^{1/3}}{P_{\rm
obs}^{1/3}}\frac{\sin
i}{c}}\times \\ \\
~~~~~~~~~~~~~~~~\displaystyle{\left(1 -\frac{v_z}{c}\cos i
\right)^{-2/3}(1-e^2)^{-1/2}}~.
\end{array}
\end{equation}

The percentage flux variation given by the previous equation
depends on the orbital parameters of the SMBHB systems, the jet
viewing angle $i$ and the jet Lorentz factor $\gamma_{\rm L}$. As
discussed above, for a typical blazar the latter quantities depend
on each other (i.e. $i\simeq 1/\gamma_{\rm L}$) while for radio
galaxies like 3C66B this is not always true.

\begin{figure}[htbp]
\begin{center}
\vspace{7.6cm} \includegraphics{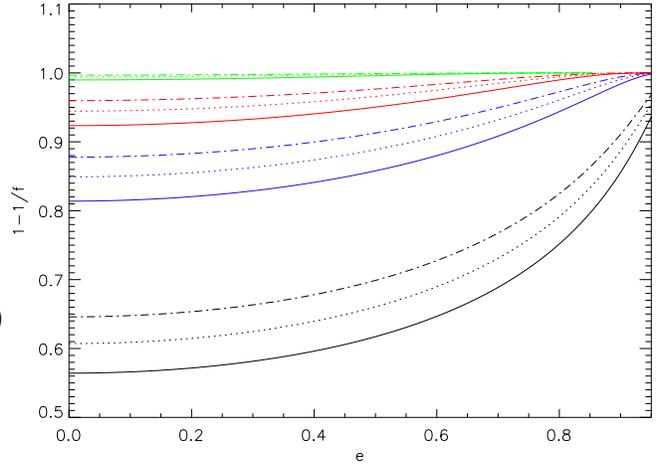} \caption{The expected percentage flux
variation of the $X$-ray and/or $\gamma$-ray light curves as a
function of the orbit eccentricity. The curves have been obtained
assuming, for the black hole maximum semi-major axis, $a\simeq
5.4\times 10^{16} (1+q)/\eta$ cm with $\eta\simeq 4$. Black, blue,
red and green lines correspond to Lorentz factors $\gamma_{\rm L}$
of 2, 4, 6 and 10, respectively. Solid, dotted and dashed lines
correspond to power law indices $\alpha=1, 1.5$ and $2$,
respectively. Here we have also assumed
 $i\simeq 1/\gamma_{\rm L}$.} \label{figure3}
\end{center}
\end{figure}

In Fig. \ref{figure3} we show the expected percentage flux
variation between the maximum and minimum signal as a function of
the orbital eccentricity, for different values of both the Lorentz
factor $\gamma _{\rm L}$ and the power law spectral index
$\alpha$. Here we have considered the galaxy as a typical blazar,
thus setting $i\simeq 1/\gamma_{\rm L}$ in all the previous
equations.

Since we expect that future radio observations at frequencies
higher than 8.4 GHz will improve the mapping of the SMBHB orbit in
3C66B, we consider, as an example, the orbit semi-major axis  to
be $a\simeq 5.4\times 10^{16} (1+q)/\eta$ cm, being $\eta\simeq
4$.

Note that in this case, using the orbit semi-major axis obtained
by the 8.4 GHz radio map (i.e. $a_{max}\simeq 5.4\times 10^{16}
(1+q)$ cm), Eq. (\ref{fluxratio}) will be satisfied only for small
Lorentz factor values $\gamma_{\rm L} \ut< 2$. In this case, the
expected flux ratio $f$ is high enough to make $1-1/f$ approach
unity for any value of the eccentricity $e$ (and therefore always
imply $f \ut> 100$).

\begin{figure*}[htbp]
\vspace{0.2cm}
\begin{center}
$\begin{array}{c@{\hspace{0.4in}}c} \epsfxsize=3.25in
\epsffile{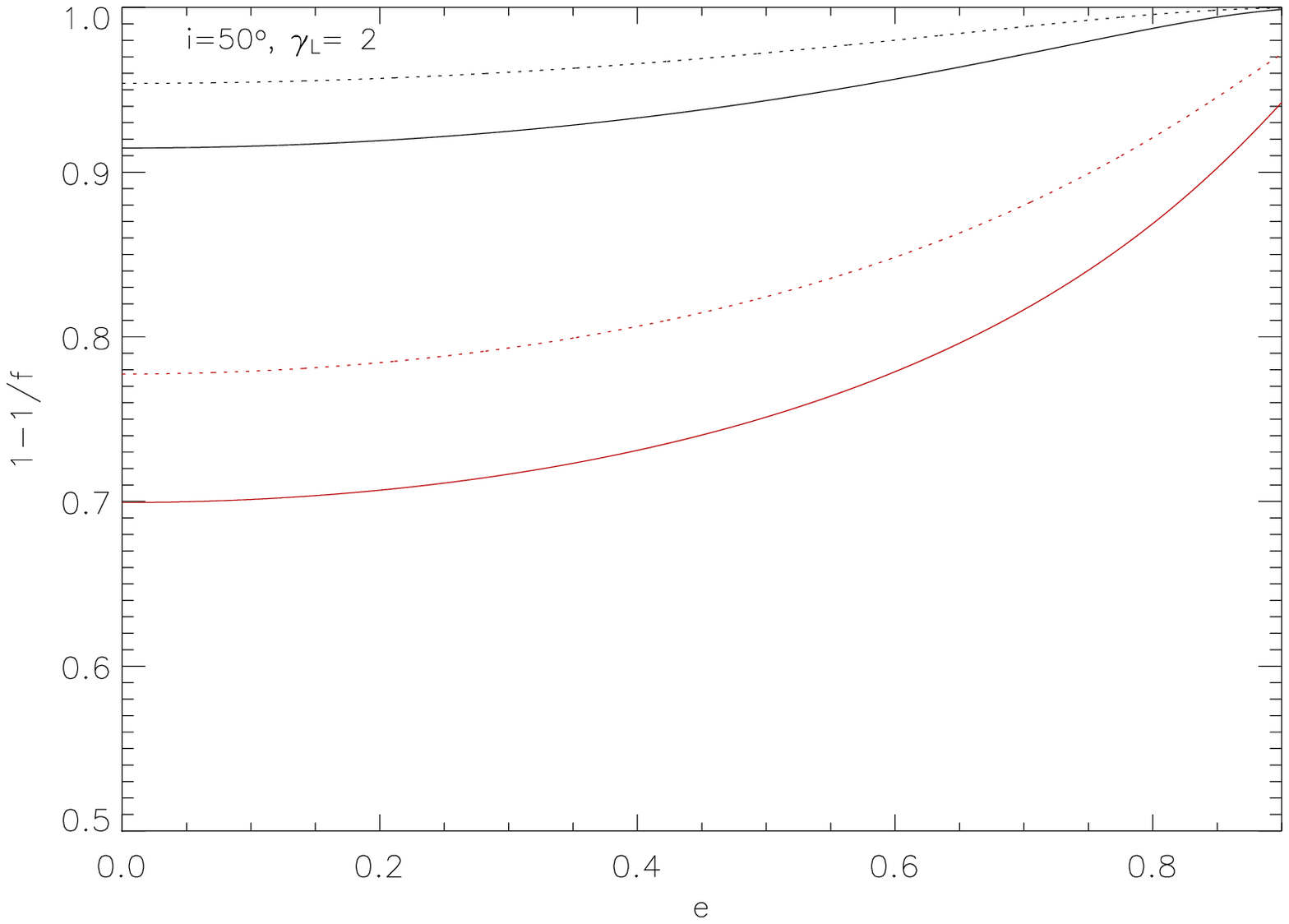} & \epsfxsize=3.25in
\epsffile{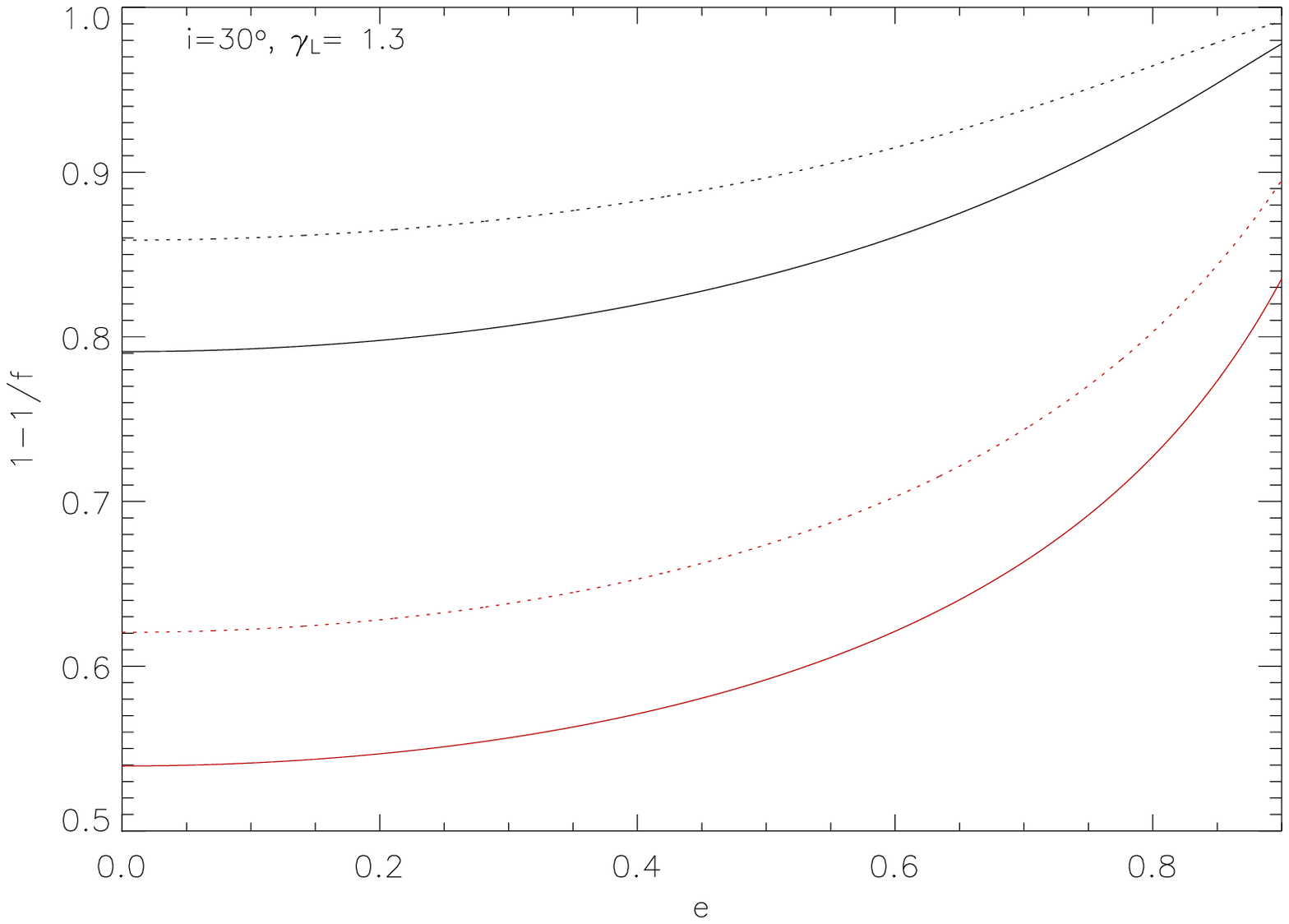}
\\ [0.cm]
\mbox{\bf a)} & \mbox{\bf b)}
\end{array}$
\end{center}
\caption{The expected percentage flux variation of the $X$-ray
and/or $\gamma$-ray light curves as a function of the orbit
eccentricity. The curves were obtained assuming, for the black
hole semi-major axis, $a \simeq 5.4\times 10^{16} (1+q)/2$ cm
(black lines) and $a \simeq 5.4\times 10^{16} (1+q)/4$ cm (red
lines). Solid and dotted lines correspond to power law index
$\alpha=1$ and $2$, respectively. Here we are considering a more
realistic geometry for the 3C66B galaxy jets.}
\label{figurefluxnew1}
\end{figure*}

Let us now consider again the most realistic scenario according to
which the jet inclination angle is $i\ut< 53\degr$ (Hardcastle et
al. \citeyear{hardcastle1996}). In this case, for each possible
value of the inclination angle $i$ there exists a different set of
allowed values of the Lorentz factor $\gamma_{\rm L}$ (see
discussion above). Hence, using Eq. (\ref{fluxratio}) we can
evaluate the expected flux percentage variation for different jet
inclination angles $i$. This is shown in Figs.
\ref{figurefluxnew1} a) and b) where we plot the expected
percentage flux for two selected values of the jet inclination
angle and $\gamma_{\rm L}$ ($i=50\degr$ and $\gamma_{\rm L}=2$ in
Fig. \ref{figurefluxnew1}a, $i=30\degr$ and $\gamma_{\rm L}=1.3$
in Fig. \ref{figurefluxnew1}b).

As one can see, $X$-ray and/or $\gamma$-ray observations towards
3C66B may make it possible to detect a possible periodic behavior
in the light curves which can be directly used to fix the Lorentz
factor $\gamma_{\rm L}$ (see Fig. \ref{figure2}). In addition, the
detection of the signal flux ratio $f$ will allow to determine the
orbit eccentricity $e$ of the SMBHB possibly hosted by the radio
galaxy 3C66B at least if the percentage flux variation is not
close to unity, and also give information about the jet
inclination angle $i$.

Note also that radio observations of the nuclear region of 3C66B
have shown a strong flux variability (up to $35$ per cent at high
frequencies) on a time scale of months (Hardcastle et al.
\citeyear{hardcastle1996}) which might be due to the Doppler
modulation effect of the jet and counterjet motion towards or away
from the observer.

\section{Discussion and conclusions}

As stated in the previous section, if 3C66B really hosts an SMBHB
in its center, a periodic behavior of the light curve in the
$X$-ray and/or $\gamma$-ray bands and a certain flux ratio $f$ are
expected. A measure of $P_{\rm obs}$ and $f$ would allow getting
both the jet Lorentz factor $\gamma _{\rm L}$ and the system orbit
eccentricity $e$ through Fig. \ref{figure2} and Fig.
\ref{figure3}.

However, since the curves in Fig. \ref{figure3} do not depend on
the mass ratio $q$, also measuring $f$ does not make it possible
to obtain a complete definition of the orbital parameters of the
SMBHB. Thus, a third observation, i.e. the gravitational wave
emission, is necessary to extract the value of $q$.

First, note that the emission of gravitational radiation makes the
SMBHB orbit to shrink within the merging time-scale (Peters
\citeyear{peters}, Fitchett \citeyear{fitchett})
\begin{equation}
T\simeq\frac{167\tau c^5}{Mm(M+m)}\int _0 ^{e_0}
\frac{\left(1+\frac{121e^2}{304}\right)^{1181/2299}e^{29/10}}{(1-e^2)^{3/2}}~de~,
\label{timescale}
\end{equation}
where $\tau$ is a constant that depends on the initial orbit
semi-major axis $a_0$ and eccentricity $e_0$.

\begin{figure}[htbp]
\begin{center}
\vspace{7.6cm} \includegraphics{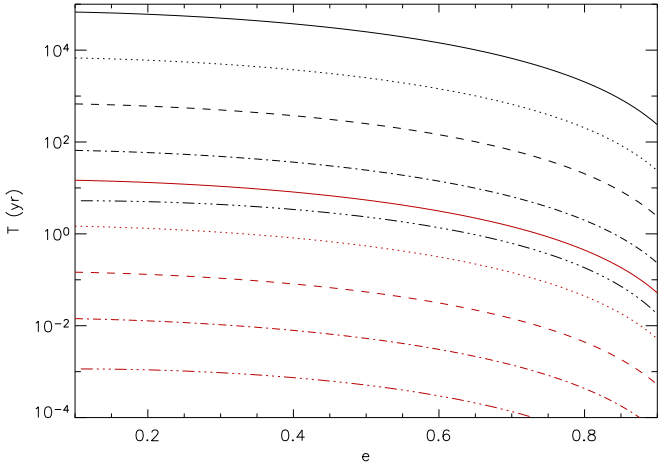} \caption{The coalescing time scale $T$ as
a function of the true orbit eccentricity $e$. Black and red lines
correspond to the orbital maximum semi-major axis $a\simeq
5.4\times 10^{16} (1+q)$ cm and $a\simeq 2.9\times 10^{17} (1+q)$
cm, respectively. Solid, dotted, dashed, dot-dashed and
two-dot-dashed lines are for $q=10^{-5}$, $10^{-4}$, $10^{-3}$,
$10^{-2}$, $10^{-1}$, respectively.} \label{figure5}
\end{center}
\end{figure}

In Fig. \ref{figure5} the coalescing time scale $T$ for different
values of the mass ratio $q$ (different lines) and two values of
the semi-major axis $a$ (different colors) is shown as a function
of the orbit eccentricity $e$. Models corresponding to the higher
value of $a$ (red lines) are clearly  unlikely since the
corresponding coalescing time-scale is shorter than $10^2$ yrs. On
the contrary, models with the smaller semi-major axis
(corresponding to that derived by using the 8.4 GHz map) and
$q\ut< 10^{-3}$ always have coalescing time-scales longer than
$\sim 10^2$ yrs and may therefore be considered acceptable. Note
also that, as previously stated, since future radio observations
at frequencies higher than 8.4 GHz will allow better constraint of
the semi-major axis of the SMBHB system the true coalescing
time-scale of the massive black hole binary systems may be longer
than that quoted here.

\begin{figure}[htbp]
\begin{center}
\vspace{7.6cm} \includegraphics{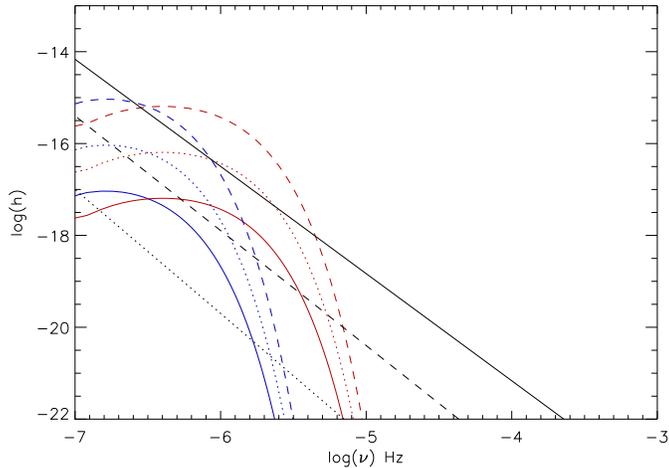} \caption{The expected gravitational wave
spectrum for the SMBHB in 3C66B. Black, blue and red lines are for
$q=10^{-5}$,$10^{-4}$, $10^{-3}$, respectively. Solid and dotted
lines indicate eccentricity values of $e=0.7$ and $e=0.9$. Here we
have assumed the semi-major axis to be $a\simeq 5.4\times 10^{16}
(1+q)$ cm. Solid, dashed and dotted straight lines show the
planned sensitivity thresholds of LISA, ASTROD1 and ASTROD2
instruments. Here we have assumed an integration time of $5$ yrs.}
\label{figure6}
\end{center}
\end{figure}

The SMBHB mass ratio $q$ may be determined if one could observe
the gravitational wave spectrum. In Fig. \ref{figure6} we present
the expected gravitational wave spectrum for different values of
the mass ratio $q$ and orbital eccentricity $e$. Straight lines
correspond to the detection thresholds of the next generation of
space-based interferometers LISA (Reinhard \citeyear{Rein00}) and
ASTROD (Wu et al. \citeyear{Wu00} and Ni \citeyear{Ni02}).

Thus, future $X$-ray and/or $\gamma$-ray observations may allow to
obtain information about both the bulk motion of the jet
($\gamma_{\rm L}$) and its precession and, consequently, confirm
the SMBHB model for the radio galaxy 3C66B. These measurements may
also allow to define the semi-major axis $a$ and estimate the
orbital eccentricity $e$ of the binary system as follows from
Figs. \ref{figure2} and Fig. \ref{figure3}. Moreover, the mass
ratio $q$ may be attained from the observation of the emitted
gravitational wave spectrum by using next generation space-based
interferometers.

Note that recently a different method for constraining the
properties of the SMBHB in 3C66B has been proposed (Jenet et al.
\citeyear{jenet}). The main idea is to search for gravitational
wave emission from the SMBHB system by using available timing data
($7$ yrs up to now) from the radio pulsar PSR B1855+09 (Kaspi et
al. \citeyear{kaspi}). In fact, gravitational waves will induce
periodic oscillations in the arrival times of the individual
pulses of the pulsar. Numerical simulations performed for
different parameters have shown that the proposed method can be
used to solve the parameter degeneracy of the SMBHB in 3C66B only
for extremely small orbital eccentricity values ($e< 0.03$). In
the other cases the method can not give a definitive answer using
the current data (Jenet et al. \citeyear{jenet}).

\begin{acknowledgements}
One of us (A.A.N.) would like to thank B.M.T. Maiolo for reading
the manuscript and A.F. Zakharov for interesting discussion. We
also would like to thank the anonymous referee for the
suggestions.
\end{acknowledgements}

\end{document}